\documentstyle[12pt]{article}
\begin{document}
\newcommand{\mup}{{\mu}^{\prime}}
\newcommand{\nup}{{\nu}^{\prime}}
\newcommand{\muz}{{\mu}^{\prime \prime}}
\newcommand{\nuz}{{\nu}^{\prime \prime}}
\newcommand{\z}{\prime \prime}
\newcommand{\p}{\prime}
\begin{titlepage}
\vspace{-10mm}
\begin{flushright}
             September 2001\\ IUT-Phys/01-11\\              
\end{flushright}
\vspace{12pt}
\begin{center}
\begin{large}             
{\bf Constraint structure \\ in  modified Faddeev-Jackiw method}\\
\end{large}
\vspace{5mm}
{\bf A. Shirzad 
\footnote{e-mail: shirzad@theory.ipm.ac.ir}}
{\bf , M. Mojiri
\footnote{e-mail:mojiri@sepahan.iut.ac.ir}}\\
\vspace{12pt} 
{\it Institute for Studies in Theoretical Physics and Mathematics\\ 
P. O. Box: 5746, Tehran, 19395, IRAN,\\
Department of  Physics, Isfahan University of Technology \\ 
Isfahan,  IRAN.} 

\vspace{0.3cm}
\end{center}
\abstract{We show that in modified Faddeev-Jackiw formalism, first and second 
class constraints appear at each level, and the whole constraint structure is 
in exact correspondence with level by level method of Dirac formalism.
\ 
\vfill}
\end{titlepage}

The standard method for analysis of the constrained systems is known to be 
the Dirac formalism \cite{Dirac}. Faddeev and Jackiw \cite{FJ}, however,
have proposed an alternative method (FJ formalism). Their approach is based on solving the 
constraints at each level, and using the Darboux theorem \cite{Dar} to   
find a smaller phase space together with a number of additional coordinates.
The equations of motion of these additional coordinates will survive for new
constraints, and the procedure will be repeated.

In a newer approach to FJ formalism \cite{Sym1,Sym2}, called 
{\it symplectic analysis } or {\it modified Faddeev-Jackiw } (MFJ) formalism,
instead of solving the constraints, one adds their time derivatives to the
Lagrangian and considers the corresponding Lagrange multipliers as additional
coordinates.
In this way, constraints would be introduced in the kinetic part of the
Lagrangian, rather than the potential. There are some efforts to show the
equivalence of FJ or MFJ method with Dirac method. In \cite{GP}, using a
special representation of constrains, it is shown that principally first
and second class constraints do appear at a typical step of FJ formalism.
However, within the Dirac formalism there exists a well-established
constraint structure, such that at each level of consistency, constraints
divide to first and second class ones \cite{BGP}. Then the consistency
of the second class constraints determines some of Lagrange multipliers,
while the consistency of first class ones leads to constraints of the
next level.

Such a constraint structure is not known in FJ or FJM formalism.
We show in this paper that the constraint structure of MFJ formalism  
emerges in the same way as in Dirac formalism. However, this structure,
which is exactly similar to that given in \cite{BGP}, is somehow hidden within
the Darboux theorem or within general statements. It should be noted that some
signals of this structure has been recognized in \cite{MONTA}, but that belong
to cases where only first or second class  constraints are present.

Suppose we are given the first order Lagrangian
\begin{equation}
L= a_i(y) \dot{y}^i-H(y) \label{a1}
\end{equation}
in which $y^i$ are coordinates not nessesarily canonical of a $2N$ dimensional
phase space.
The equations of motion read
\begin{equation}
f_{ij} {\dot{y}}^j = \partial_i H \label{a2}
\end{equation} 
where $\partial_i$ means $ \partial / \partial y^i$ and 
\begin{equation}
f_{ij} \equiv \partial_i a_j(y) - \partial_j a_i(y). \label{a3}
\end{equation}

The tensor $f_{ij}$ is called the {\it Presymplectic} tensor. We denote it in matrix
notation as $f$. If $y^i$ are chosen to be the usual canonical variables
$(q,p)$ then the term ${a_i}(y){\dot{y}}^i$ in (\ref{a1}) would be $p_i \dot{q}^i$.
In other words, in some general (non canonical) coordinates $y^i$ of the phase
space, the term ${a_i}(y){\dot{y}}^i$ has the same role as $p_i \dot{q}^i$.
Therefore, according to construction of first order Lagrangian, it is
reasonable that $f$ is not singular. In fact the constraints would be imposed
thereafter to a Lagrangian with nonsingular presymplectic tensor, as we will
see. However, if for some artificial Lagrangian, $f$ is singular with null
eigenvectors $v^i_a(y)$, then constrains $ \Phi_a(y) = v^i_a(y) \partial _i H(y)$
would result from equations of motion (\ref{a2}). In this case, one can treat
with the constraints in the same manner as we will do in the following, and
there in  no serious difference. So, from now on we assume that the presymplectic
tensor $f$ is not singular. Let denote components of $f^{-1}$ as $f^{ij}$.
Then equations (\ref{a2}) read
\begin{equation}
{\dot{y}}^i = f^{ij} \partial_j H. \label{a4}
\end{equation} 

The antisymmetric tensor $f^{ij}$ defines the Poisson brackets of
phase space coordinates:
\begin{equation}
\left\{ y^i,y^j \right\} =f^{ij}. \label{a5}
\end{equation}

Assuming that
\begin{equation}
\left\{ F(y),G(y)\right\}=\partial_i F\partial_j G \left\{ y^i,y^j \right\} 
\label{b1}
\end{equation}
the Poisson bracket of two arbitrary functions $F(y),G(y)$ can be obtained
as
\begin{equation}
\left\{ F(y),G(y) \right\}=\partial_i F f^{ij}\partial_j G. \label{b2}
\end{equation}

Suppose we want to impose the primary constraints 
$\Phi^{(0)}_{\mu}(\mu=1,...,M)$ to the system given by Lagrangian (\ref{a1}).
This can be done by considering the Lagrangian
\begin{equation}
L^{\p}= a_i(y) \dot{y}^i-H(y)-v^\mu \Phi^{(0)}_\mu \label{a6}
\end{equation}
with $v_\mu$ as additional coordinates. Hence the presymplectic tensor
$f$ should be replaced by the $(2N+M) \times (2N+M)$ singular tensor
\begin{equation}
f^\prime = \left( \begin{array}{c|c} f &0 \\ \hline 0&0 \end{array} \right). 
\label{a7}
\end{equation}
Multiplying both sides of (\ref{a2}) with the null eigenvevtors of (\ref{a7})
restores the constraints $\Phi^{(0)}_{\mu}$. Therefore the additional term 
$v^{\mu}\Phi^{(0)}_{\mu}$ in (\ref{a6}) has no new result. However, consistency
of the constraints during time imposes new constraints on system,
i.e. $d \Phi^{0}_{\mu} / dt $ should vanish. Adding the term $\eta^{\mu}\dot{\Phi^{0}_{\mu}}$ 
to $L$, the modified Lagrangian would be
\begin{equation}
L^{(1)}=\left(a_i-\eta^{\mu}A_{\mu i} \right) {\dot{y}}^i-H(y), \label{a8}
\end{equation}
where
\begin{equation}
A_{\mu i} \equiv \partial_i \Phi^{(0)}_{\mu}. \label{a9}
\end{equation}
It should be noted that , as shown in \cite{GP}, solving the constraints and 
considering the equations of motion for the non canonical coordinates are
equivalent to considering the consistency of constraints $\Phi^{(0)}_{\mu}$.

Considering $Y \equiv (y^i,\eta^{\mu})$ as coordinates, the symplectic tensor 
$f^{(1)}$ reads
\begin{equation}
f^{(1)} = \left( \begin{array}{c|c}f & A \\ \hline - \tilde{A} & 0 \end{array} \right) 
 \label{a10}
\end{equation}
where the elements of $(2N \times M)$ matrix A are defined in (\ref{a9}) and 
$\tilde{A}$ is the transposed matrix. The equations of motion in matrix 
notation then reads 
\begin{equation}
f^{(1)} \dot{Y}= \partial H. \label{b3}
\end{equation}
                                                            
Using operations that keep the determinant invariant, it is easy to show that 
\begin{eqnarray}
\nonumber\det f^{(1)}&=&\det \left( \begin{array}{c|c} f& A \\ \hline 0 &
\tilde{A}f^{-1}A 
\end{array} \right) \nonumber \\ {} &=&(\det f)(\det \tilde{A} f^{-1} A).
\end{eqnarray}
Assuming that $ \det f \neq 0 $, $ f^{(1)} $ would be singular if $ C \equiv \tilde{A} f^{-1} A $ 
is singular. Using (\ref{a5}) and (\ref{a9}) we have 
\begin{equation}
C_{\mu \nu}=\left\{\Phi^{(0)}_{\mu},\Phi^{(0)}_{\nu} \right\}. \label{a11}
\end{equation}
Therefore, the singularity disappears if $C_{\mu \nu}$ is invertible
(this is the case considered in \cite{MONTA}).
In Dirac terminology this means that all primary constrains are second class. 
In this case ${f^{(1)}}^{-1}$ defines a new bracket in the same way as in
(\ref{b2}). It can be shown that: 
\begin{equation}
{f^{(1)}}^{-1} = \left( \begin{array}{c|c}f^{-1}-f^{-1} A C^{-1} \tilde{A} f^{-1}
& -f^{-1} A C^{-1} \\ \hline C^{-1} \tilde{A} f^{-1} & C^{-1} \end{array} \right). 
\label{b4}
\end{equation}
Using (\ref{b4}) one can easily see that the new bracket between two 
functions of the original phase space is the well-known Dirac bracket:
\begin{eqnarray}
\left\{ F,G \right\}_{D.B} &=& \left\{ F,G \right\} - \left\{ 
F,\Phi^{(0)}_{\mu} \right\} C^{\mu \nu} \left\{ \Phi^{(0)}_{\nu},G \right\} \label{a12}
\end{eqnarray}
where $C^{\mu \nu}$ is the inverse of $C_{\mu \nu}$.

Now suppose that $ C_{\mu \nu}=\left\{ \Phi^{(0)}_{\mu},\Phi^{(0)}_{\nu} \right\} $ 
is singular. As in the framework of Dirac formalism \cite{BGP}, we can assume that
the constraints $\Phi^{(0)}_{\mu}$ divide to $M^{\p}$ first and $ M^{\z}$ 
second class constraints $ \Phi_{\mup}^{(0)}$ and $\Phi_{\muz}^{(0)}$
respectively $(M^{\p}+M^{\z}=M)$ with the following property 
\begin{equation}
\begin{array}{lcr}
\left\{ \Phi^{(0)}_{\mup},\Phi^{(0)}_{\nup} \right\} \approx 0 & { }\\
\left\{ \Phi^{(0)}_{\mup},\Phi^{(0)}_{\nuz} \right\} \approx 0 & { } \\
\left\{ \Phi^{(0)}_{\muz},\Phi^{(0)}_{\nuz} \right\} 
\approx C_{\muz \nuz} & { }& \det{C_{\muz \nuz}} \neq 0 \label{a13}
\end{array}
\end{equation}
where $ \approx $ means weak equality, i.e. equality on the surface of constraints
known already (here, Primary constraints). Defining
\begin{equation}
\begin{array}{lr}
A_{\mup i}= \partial_i \Phi^{(0)}_{\mup} & { } \\
A_{\muz i}= \partial_i \Phi^{(0)}_{\muz} \label{a14}
\end{array}
\end{equation}
the tensor $f^{(1)}$ reads
\begin{equation}                
{f^{(1)}} = \left( \begin{array}{c|c|c} f & A^ {\z} & A^ {\p} \\ \hline
-{\tilde{A}}^ {\z} & 0 & 0 \\ \hline -{\tilde{A}}^ {\p} & 0 & 0 \end{array} \right). 
\label{a15}
\end{equation}
Using (\ref{a13}) and (\ref{a11}) it is easy to show that the rows of
\begin{equation}
\left({\tilde{A}}^ {\p}f^{(-1)},0,1 \right) \label{a16}
\end{equation}
which is an $ M^{\p} \times \left( 2N + M^{\z} + M^{\p}
\right) $ matrix, are null eigenvectors of $ f^{(1)} $. Multiplying both sides
of (\ref{b3}) with each row of (\ref{a16}), say row $\mup$, gives the new constraint 
\begin{equation}
\Phi^{(1)} _ {\mup} \equiv \left\{ \Phi^{(0)}_{\mup},H \right\}=0. \label{a17}
\end{equation}
This is exactly the same result as obtained in Dirac formalism \cite{BGP},
i.e. the consistency of first calls constraints at each level gives the constraints 
of the next level. 

On the other hand, note that the $f^{(1)}$ matrix in (\ref{a15}) 
has an invertible sub-block
\begin{equation}              
f^{(1)}_{\rm inv} = \left( \begin{array}{c|c} f & A^ {\z} \\ \hline - {\tilde{A}}^ {\z} 
& 0 \end{array} \right) \label{a18}
\end{equation}
The inverse of (\ref{a18}) is similar to (\ref{b4}). Dividing Lagrange
multipliers $\eta^\mu$ to $\eta^{\mup}$ and $\eta^{\muz}$ corresponding
to first and second class constraint respectively, one can use the inverse of (\ref{a18})
to find ${\dot{\eta}}^{\muz}$ from (\ref{b3}). The result is
\begin{equation}
\dot{\eta}^{\muz}=-C^{\muz \nuz} \left\{\Phi^{(0)}_{\nuz}, H \right\}, \label{a19}
\end{equation}
where $C^{{\muz} {\nuz}}$ is the inverse of
\begin{equation}
C_{{\muz} {\nuz}}= \left\{\Phi^{(0)}_{\muz},\Phi^{(0)}_{\nuz} \right\}. 
\label{a20}
\end{equation}
Adding a total derivative, one can replace the term 
$ \eta ^{\muz} \dot{\Phi}^{(0)}_{\muz} $ in Lagrangian (\ref{a8}) with 
$- \dot{\eta} ^{\muz} {\Phi}^{(0)}_{\muz} $. Then from (\ref{a19}),
one observes that the replacement
\begin{equation}
H^{(1)}=H^{(0)}-\left\{H^{(0)},\Phi^{(0)} _ {\muz} \right\} C^{{\muz} {\nuz}} 
\Phi^{(0)}_{\nuz} \label{a21}
\end{equation}
suffices to omit second class constraints $ \Phi^{(0)}_{\muz} $ in the remaining 
and to work only with first calls constraints $ \Phi^{(0)}_{\mup}$.
Hence, before going to next level, we assume that our Lagrangian is
\begin{equation}
L^{(1)}=\left(a_i- \eta^{(0) \mu} \frac{\partial \Phi^{(0)}_ \mu}{\partial y^i} 
\right){\dot{y}}^i - H^{(1)}(y) \label {a22}
\end{equation}
where all primary constraints $\Phi^{(0)}_{\mu}$ are first class. (These are in fact the previous
$ \Phi^{(0)}_{\mup} $, where for simplicity we have omitted the primes.)
The corresponding symplectic tensor reads 
\begin{equation}
f^{(1)} = \left( \begin{array}{c|c} f & A^{(0)} \\ \hline  -{\tilde{A}}^{(0)} 
& 0 \end{array} \right) 
\label{a23}
\end{equation}
where $ A^{(0)} \equiv \partial _i \Phi^{(0)}_{\mu} $ stands for gradient of
primary constraints which, as stated, are assumed to be first class so far.

Now let's go to the level step. This aim would be achieved by adding the consistency
term $ -\eta^{(1) \mu} {\dot{\Phi}}^{(0)}_\mu $ to the Lagrangian (\ref{a22})
to give
\begin{equation}
L^{(2)}=\left(a_i- \eta^{(0) \mu} \frac{\partial \Phi^{(0)}_ \mu}{\partial y^i} 
- \eta^{(1) \mu} \frac{\partial \Phi^{(1)}_ \mu}{\partial y^i} \right)
{\dot{y}}^i - H^{(1)}(y) \label {a24}
\end{equation}
Taking $ \eta^{(1)}_{\mu} $ as new variables, we have 
\begin{equation}                
{f^{(2)}} = \left( \begin{array}{c|c|c} f & A^ {(0)} & A^ {(1)} \\ \hline
-{\tilde{A}}^ {(0)} & 0 & 0 \\ \hline -{\tilde{(A)}}^ {(1)} & 0 & 0 \end{array} \right) 
\label{a25}
\end{equation}
As we will see (for an arbitrary level), $\det f^{(2)}$ vanishes if the matrix
$ C^{(1)}_{{\mu} {\nu}}= \left\{\Phi^{(0)}_{\mu},\Phi^{(1)}_{\nu} \right\} $
is singular, and vise versa. If $ \det C^{(1)} \neq 0 $ then the singularity
disappears at all. This means that the system is completely second class.
If, however, $ C^{(1)} $ is singular, then one can in principle divide
$ \Phi^{(1)}_{\mu} $ to $ \Phi^{(1)}_{\mup} $ and $ \Phi^{(1)}_{\muz} $, and
$ \Phi^{(0)}_{\mu} $ to $ \Phi^{(0)}_{\mup} $ and $ \Phi^{(0)}_{\muz} $
in such a way that
\begin{equation}
\begin{array}{lcr}
\left\{ \Phi^{(1)}_{\mup},\Phi^{(0)}_{\mu} \right\} \approx 0 & { }\\
\left\{ \Phi^{(1)}_{\muz},\Phi^{(0)}_{\nuz} \right\} \approx C^{(1)}_{\muz \nuz} 
&{ } & \det{C^{(1)}_{\muz \nuz}} \neq 0 \label{a26}
\end{array}
\end{equation}
Rearranging $f^{(2)} $ one can find the invartible sub-block:
\begin{equation}                 
f^{(2)}_{\rm inv} = \left( \begin{array}{c|c|c} f & A^{(0) {\z}} & A^{(1) {\z}} \\ \hline
-{\tilde{A}}^ {(0) {\z}} & 0 & 0 \\ \hline -{\tilde{A}}^ {(1) {\z}} & 0 & 0 \end{array} \right). 
\label{a27}
\end{equation}
Using inverse of (\ref{a27}), $ {\dot{\eta}}^{(0)}_{\muz} $ and
$ {\dot{\eta}}^{(1)}_{\muz} $ can be derived from the equation of
motion (\ref{b3}). This leads to defining a newer Hamiltonian $H^{(2)}$ whose
explicit form is not important. Moreover, multiplying the equations of motion
with the null eigenvectors of
\begin{equation}
\left({\tilde{A}}^ {(1) {\p}}f^{(-1)},0,0,0,1 \right) \label{a29}
\end{equation}
gives the third level constraints 
\begin{equation}
\Phi^{(2)} _ {\mup} = \left\{ \Phi^{(1)}_{\mup},H^{(1)} \right\}. 
\label{a30}
\end{equation}

To complete the discussion, and see some more details, let us consider  
the $n$-th step.

Suppose
\begin{equation}
\Phi^{(n)} _ {\mu} = \left\{ \Phi^{(n-1)}_{\mu},H^{(n-1)} \right\}
\label{b8}
\end{equation}
are $n$-th level constraints;
\begin{equation}
L^{(n)}=\left(a_i- \sum^{n-1}_{k=1} \eta^{(k) \mu} \frac
{\partial \Phi^{(k)}_ \mu}{\partial y^i} \right){\dot{y}}^i - H^{n}(y) 
\label {a31}
\end{equation}
is the $n$-th level Lagrangian and 
\begin{equation}                 
f^{(n)} = \left( \begin{array}{c|ccc} f & A^{(0)} & \cdots & A^{(n)} 
\\ \hline -{\tilde{A}}^{(0)} & 0 & \cdots & 0 \\ \vdots & \vdots & { }
& \vdots \\ -{\tilde{A}}^ {(n)} & 0 & \cdots & 0 \end{array} \right) \label{a32}
\end{equation}
is the $n$-th level symplectic tensor, where   
\begin{equation}
A^k_{\mu i}=\partial_i \Phi^{(k)}_{\mu}. \label{a33} 
\end{equation}
Omitting second class constraints up to level $(n-1)$, we have assumed that
constraints $ \Phi^{(0)}_{\mu},\Phi^{(1)}_{\mu} \cdots \Phi^{(n-1)}_{\mu} $
are first class.
Using operations that leave the determinant invariant, one can show that
\begin{equation}                 
\det f^{(n)} =(\det f).\det \left( \begin{array}{ccc} P^{01} & \cdots & P^{0n}  
\\ \vdots & { } & \vdots \\ P^{n0} & \cdots & P^{nn}  \end{array} \right)
\label{a34}
\end{equation}
where
\begin{equation}
P^{kl}={\tilde{A}}^k f^{(-1)} A^l= \left\{ \Phi^{(k)}, \Phi^{(l)} \right\}.
\label{a35}
\end{equation}
Using Jacobi identity and the recursion relation
\begin{equation}
\Phi^{(k)}_{\mu} \approx \left\{ \Phi^{(k-1)}_{\mu}, H \right\}
\label{a36}
\end{equation}
it is easy to show
\begin{equation}
\left\{ \Phi^{(k)}_{\mu}, \Phi^{(l)}_{\nu} \right\} \approx - \left\{ 
\Phi^{(k-1)}_{\mu}, \Phi^{(l+1)}_{\nu} \right\}. \label{a37}
\end{equation}
Since $ \Phi^{(0)}_{\mu},\Phi^{(1)}_{\mu} \cdots \Phi^{(n-1)}_{\mu} $
are assumed to be first class, using (\ref{a37}) we observe that  
\begin{equation}
\begin{array}{lccr}
P^{kl} \approx 0 & {\rm if} & { } & k+l <n \label{a38}
\end{array}
\end{equation}
In other words the matrix $ P $ has the following form 
\begin{equation}                 
P = \left( \begin{array}{ccccc} 0 & 0 & \cdots & 0 & P^{0n}  
\\ 0 & 0 & \cdots & P^{1 (n-1)}&P^{1n} \\ \vdots & \vdots & { } & \vdots &\vdots \\  
0 & P^{(n-1) 1} & \cdots & P^{(n-1) (n-1)} & P^{(n-1) n} \\
P^{n0} & P^{n1} & \cdots & P^{n (n-1)} & P^{nn }\end{array} \right). \label{a39}
\end{equation}
Then using (\ref{a37}) we can write 
\begin{equation}
\det P \propto \left(\det{P^{0n}} \right)^n. \label{a40}
\end{equation}
Therefore, $ \det f^{(n)} $  vanishes if $ \left\{ \Phi^{(0)}_{\mu},\Phi^{(n)}_{\nu} \right\} $
is singular. This is also an important observation in Dirac theory \cite{BGP}, that 
the criterion to distinguish between first and second class constraints is singularity
of their Poisson brackets with primary constraints. If $\det P^{0n} \neq 0$ then
the singularity of the theory disappears and all of the constraints would be second
class. If, however, $P^{0n}$ is singular, one should divide all sets of constraints
to first class ones: $ \Phi^{(0)}_{\mup}, \cdots \Phi^{(n)}_{\mup} $
and second class ones: $ \Phi^{(0)}_{\muz}, \cdots \Phi^{(n)}_{\muz} $.
The matrices $ A^{(0)},\cdots A^{(n)} $ also break into $ A^{(0)  \p},\cdots A^{(n) \p} $
and  $ A^{(0)  \z},\cdots A^{(n) \z} $ accordingly. Rearranging $f^{(n)}$ 
such that  $ A^{(0)  \z},\cdots A^{(n) \z} $ come first, we will find the invertible sub-block as
\begin{equation}                
f^{(n)}_{\rm inv} = \left( \begin{array}{c|ccc} f & A^ {(0) \z} & \cdots
& A^{(n) \z} \\ \hline -{\tilde{A}}^ {(0),\z} & 0 & \cdots & 0 \\
\vdots & \vdots & { } & \vdots \\ -{\tilde{A}}^ {(n) {\z}} & 0 & \cdots
& 0 \end{array} \right) \label{a41}
\end{equation}
The inverse can be found to be
\begin{equation}
{f^{(n)}_{\rm inv}}^{ -1} = \left( \begin{array}{c|c}f^{-1}-f^{-1} A C^{-1} \tilde{A} f^{-1}
& -f^{-1} A C^{-1} \\ \hline C^{-1} \tilde{A} f^{-1} & C^{-1} \end{array} \right) 
\label {a42}
\end{equation}
where $A$ denotes the matrix:
\begin{equation}
A = \left( \begin{array}{c|c|cc} A^{(0) \z} & A^{(1) \z} & \cdots & A^{(n) \z}
\end{array} \right). \label {b7}
\end{equation}
This enables us to solve the equations of motion to find
$ {\dot{\eta}}^{(0) {\z}}_{\mu}, \cdots {\dot{\eta}}^{(n) {\z}}_{\mu} $. Substituting
into  Lagrangin, the modified Hamiltonian $H^{(n+1)}$ would be derived.
Finally the null-eigenvectors of $ f^{(n)} $ (after rearranging) can be written as
\begin{equation}
({\tilde{A}}^ {(n),\p} f^{(-1)},\stackrel{2n-1}{\overbrace{0,0,\cdots ,0}},1). 
\label {a44}
\end{equation}
Multiplying the equations of motion with (\ref{a44}) finally gives the
constraints of the next level
\begin{equation}
\Phi^{(n+1)}_{\mu}=\left\{\Phi^{(n)}_{\mu}, H^{(n)} \right\}
\label {a45}
\end{equation}
As is observed, the whole thing goes on in MFJ formalism in the
same way as in Dirac formalism, i. e. at each level a number of second class
constraints $(\Phi^{(0)}_{\muz},\cdots,\Phi^{(n)}_{\muz})$ are separated,
and a set of new constraints $\Phi^{(n+1)}_\mu$ emerge.
The whole procedure will finish at level N, say, in two case: first, if
$\Phi^{(N+1)}_{\mu} \approx \left\{ \Phi^{(N)}_{\mu}, H \right\}$ vanish weakly
and next, if the symplectic matrix $f^{(N)}$ is invertible. In the first case the
system possesses gauge invariance generated by $(\Phi^{(0)}_{\mu},\cdots,
\Phi^{(N)}_{\mu})$ (assuming that we have put away second class constraints
in previous levels). In the second case the system is completely second class
and there is no gauge invariance.

As a final point, we give a few words about a seemingly contradiction between
Dirac and MFJ  methods appeared in \cite{Sym2} when discussing the model
\begin{equation}
L=\left( q_2+q_3 \right) \dot{q}_1 +q_4 \dot{q}_3 + \frac{1}{2} \left( q^2_4
- 2q_2q_3 - q^2_3 \right). \label{s1}
\end{equation}
This model was first introduced in \cite{HOJ} as an example that the equation
of motion can not be derived from a second order Lagrangian. Then considering
$q_i$ as coordinates of a configuration space, \cite{KUL} has discussed
constraint structure of the model in Dirac theory. In this way four second class
constraints would emerge and the dynamics on the constraint surface goes on via Dirac
brackets. The model then is analysed in \cite{Sym2} using MFJ method
and no constraint has been encountered.

To see what has happened, we consider the general form of a first order
Lagrangian as given by (\ref{a1}). We emphasize that in FJ method the variables
of the first order Lagrangian are viewed as coordinates of a phase space.
It is, however, possible to consider those variables as coordinates of a
configuration space. Hence, suppose we are given the Lagrangian
\begin{equation}
L=a_i(q) \dot{q}_i -V(q) \label{s2}
\end{equation}
which is obviously singular. Following Dirac method, the primary constraints
are
\begin{equation}
\Phi_i=p_i-a_i(q) \approx 0 \label{s3} 
\end{equation}
We see that dublicating the space of $q^i$ to phase space of $(q^i,p_i)$
is compensated by the same number of constraints $\Phi _i$.

Since there is no quadratic term in Lagrangian (\ref{s2}), the canonical
Hamiltonian is the same as $V(q)$ and the total Hamiltonian reads
\begin{equation}
H_T = V(q) + \lambda^i \left(p_i -a_i (q) \right) \label{s4} 
\end{equation}
where $ \lambda^i $ are Lagrangian multipliers. It is easy to see that
\begin{equation}
\left\{\Phi _i,\Phi _j \right\}=\frac{\partial a_j}{\partial q^i} - \frac{\partial a_i}{\partial q^j} \equiv f_{ij}.
\label{s5}
\end{equation}
Suppose $f$ is not singular. In the context of MFJ method, this leads to
solving equations of motion (\ref{a2}) to get
\begin{equation}
\dot{q}^i = f^{ij} \partial_j V. \label{s6} 
\end{equation}
However, in the context of Dirac theory, nonsingularity of $f$ means that the
constraints $\Phi _i$ are second class and consequently the constraint surface
given by $p_i=a_i(q)$ has a phase space structure by itself.

The consistency conditions $\dot{\Phi}_i \approx 0 $ gives
\begin{equation}
\left\{\Phi_i,H_T \right\} =- \partial_i V+f_{ij} \lambda^j \label{s7} 
\end{equation}
which determines all $ \lambda^j $ in terms of coordinates. Inserting in
(\ref{s4}) gives
\begin{equation}
H_T=V(q)+\left(p_i-a_i(q) \right) f^{ij} \partial _j V \label{s8} 
\end{equation}
The canonical equations of motion for $q^i$ give
\begin{equation}
\dot{q}^i = \frac{\partial H_T}{\partial p_i} = f^{ij} \partial_j V; 
\end{equation}
the same result as (\ref{s6}). (The equation $\dot{p}_i= - \frac
{\partial H_T}{\partial q^i}$ give time derivative of (\ref{s3}),
which is obvious.)

We see that for a regular first order Lagrangian the Dirac procedure (after
turning around) gives the same result as one can obtain directly form MFJ formalism.
It is not difficult to see that the same result is correct if the system is singular.
So the apparent contradiction between Dirac and FJ methods is not in fact a
contradiction. The basic point is considering $y^i$ (in first order Lagrangian)
as phase space or configuration space coordinates.

Concluding, we think that FJ and MFJ formalism are basically the same as Dirac
formalism formulated in the language of first order Lagrangians. From another
point of view our work here was based on extending the symplectic tensor at each
step and studying it's singularity properties. This, however, is closely similar
to method given in \cite{SHIR}, where for a second or higher order Lagrangian
the Hessian matrix is extended at each step and it's null eigenvectors are
searched for.

\vspace{3mm}
{\Large Acknowledgement}

\vspace{3mm}
We thank F. Loran for useful discussions.

\end{document}